\documentclass{article}
\usepackage[T1]{fontenc}
\usepackage[utf8]{inputenc}
\usepackage{ismir}
\usepackage{amsmath,cite,url}
\usepackage{graphicx}
\usepackage{color}

\usepackage{amsmath}
\usepackage{amssymb}
\usepackage{color}
\usepackage{xspace}

\usepackage{multirow}
\usepackage{tabularx}
\usepackage{tablefootnote}
\usepackage{booktabs}
\usepackage{makecell}
\usepackage{bbding}

\usepackage{amssymb}
\usepackage{pifont}

\usepackage{microtype}
\usepackage{enumitem}

\usepackage[table]{xcolor}
\usepackage{siunitx}
\usepackage{caption}
\usepackage{subcaption}
\usepackage{etoolbox}

\newcommand{\tool}{\textsc{AnySynth}\xspace}

\title{\tool: Zero-Shot Instrument Cloning via In-Context Learning and Asymmetric Hierarchical Guidance}

\multauthor
  {Chong Jing$^1$ \hspace{0.35cm} Junan Zhang$^1$ \hspace{0.35cm} Jing Yang$^2$ \hspace{0.35cm} Yulun Wu$^2$ \hspace{0.35cm} Fan Fan$^2$ \hspace{0.35cm} Zhizheng Wu$^1$}
{
  $^1$ Chinese University of Hong Kong, Shen Zhen\\
  $^2$ Central Media Technology Institute, Huawei\\
  {\tt\small chongjing@link.cuhk.edu.cn, junanzhang@link.cuhk.edu.cn}\\
  {\tt\small yangjing201@huawei.com, wuyulun9@huawei.com, fanfan1@huawei.com}\\
  {\tt\small wuzhizheng@cuhk.edu.cn}
}

\def\authorname{C. Jing, J. Zhang, J. Yang, Y. Wu, F. Fan, and Z. Wu}

\usepackage[bookmarks=false,pdfauthor={\authorname},pdfsubject={\pdfsubject},hidelinks]{hyperref}

\begin{document}

\maketitle

\begin{abstract}
Zero-shot instrument cloning aims to render an arbitrary \texttt{[Target MIDI]} sequence with the acoustic identity of an unseen instrument given only a short \texttt{[Reference Audio, Reference MIDI]} pair. Existing methods rely on pre-trained embeddings (e.g., CLAP) that compress the reference audio into a fixed-length vector, discarding fine-grained acoustic cues essential for faithful timbre reconstruction. We present \tool, an embedding-free neural synthesizer based on in-context flow matching. By conditioning a Diffusion Transformer (DiT) directly on the uncompressed reference audio and target MIDI, our model allows self-attention to dynamically retrieve acoustic details at generation time. Experiments show that \tool outperforms embedding-based and auto-regressive baselines in audio quality, timbre similarity, and melody adherence. Notably, the model exhibits \textit{prompt-length scaling}: longer reference prompts yield steadily better timbre fidelity, a property absent in embedding-based systems. To optimize controllability, we further propose \textit{Asymmetric Hierarchical CFG}, which structurally decouples MIDI and reference-timbre guidance based on their natural semantic-acoustic dependency. This asymmetric formulation avoids gradient conflicts and improves both note accuracy and timbre fidelity, pushing the boundary of expressive, zero-shot instrument cloning. Demo audios are available at \url{https://anysynth-demo.github.io/}
\end{abstract}

\section{Introduction}\label{sec:introduction}

Recent autoregressive and diffusion-based models have substantially improved music audio generation~\cite{copet2023simple, agostinelli2023musiclm,huang2023noise2music,yuan2025yue,prajwal2024musicflow,zhang2026vevo2,zhang2025anyaccomp,gong2025ace,gong2026ace,yang2025songbloom,lei2025levo,yang2026heartmula}. As synthesis quality increases, the next challenge is fine-grained controllability: rather than generating plausible audio in coarse granularity (e.g., conditioned on text like \textit{"An r\&b electric piano solo."}), musicians require systems capable of rendering a desired note sequence with a specific instrument identity. This motivates the task of \textbf{zero-shot instrument cloning}, where an arbitrary \texttt{[Target MIDI]} must be synthesized using the acoustic identity extracted from a short, aligned \texttt{[Reference Audio, Reference MIDI]} pair~\cite{demerle2024combining,yang2025flowsynth,kim2025tokensynth,liu2026nclmctt}. 

Existing state-of-the-art (SOTA) solutions approach this instrument cloning task by disentangling musical "structure" (pitch and velocity) from "timbre". Whether built upon codec language models (e.g., TokenSynth~\cite{kim2025tokensynth}) or diffusion models (e.g., Control-Transfer-Diffusion~\cite{demerle2024combining}, FlowSynth~\cite{yang2025flowsynth}), these models share a critical dependency: they heavily rely on a pre-trained semantic encoder such as CLAP or MuLan~\cite{wu2023clap,huang2022mulan} to extract a fixed-length timbre embedding. While architecturally convenient, this design introduces a severe \textbf{semantic embedding bottleneck}. Retrieval-oriented audio encoders are trained to preserve coarse semantic alignment, inherently discarding the low-level acoustic details -- such as short transients, nuanced performance artifacts, and specific recording characteristics -- required for faithful sound reconstruction. 
Consequently, the upper bound of the synthesized timbre's fidelity is severely constrained by the representation capacity of the embedding. Some recent systems reduce this dependency on retrieval embeddings~\cite{liu2026nclmctt}, yet the central challenge remains the same: \textit{How to preserve fine-grained acoustic information while still following a new target note sequence?}

To fundamentally bypass this information bottleneck, we introduce \tool, dropping explicit timbre embeddings and reformulating zero-shot instrument cloning as an \textbf{In-Context Learning (ICL)} audio generation problem~\cite{chen2025valle,tang2025midivalle,le2023voicebox,zhang2025anyenhance,chen2025f5,wang2025maskgct,wang2025metis}. By directly feeding the reference audio as a prompt and combining Flow Matching with a Diffusion Transformer (DiT) architecture, \tool operates in a continuous feature space where self-attention mechanisms interact directly with the \textit{uncompressed} reference context. Instead of forcing the model to memorize a compressed representation, \tool learns to dynamically attend to and extract raw acoustic details from the \texttt{[Reference Audio, Reference MIDI]} pair to accurately render the \texttt{[Target MIDI]}.

\begin{figure*}[!htbp]
  \centering
  \includegraphics[width=0.9\linewidth]{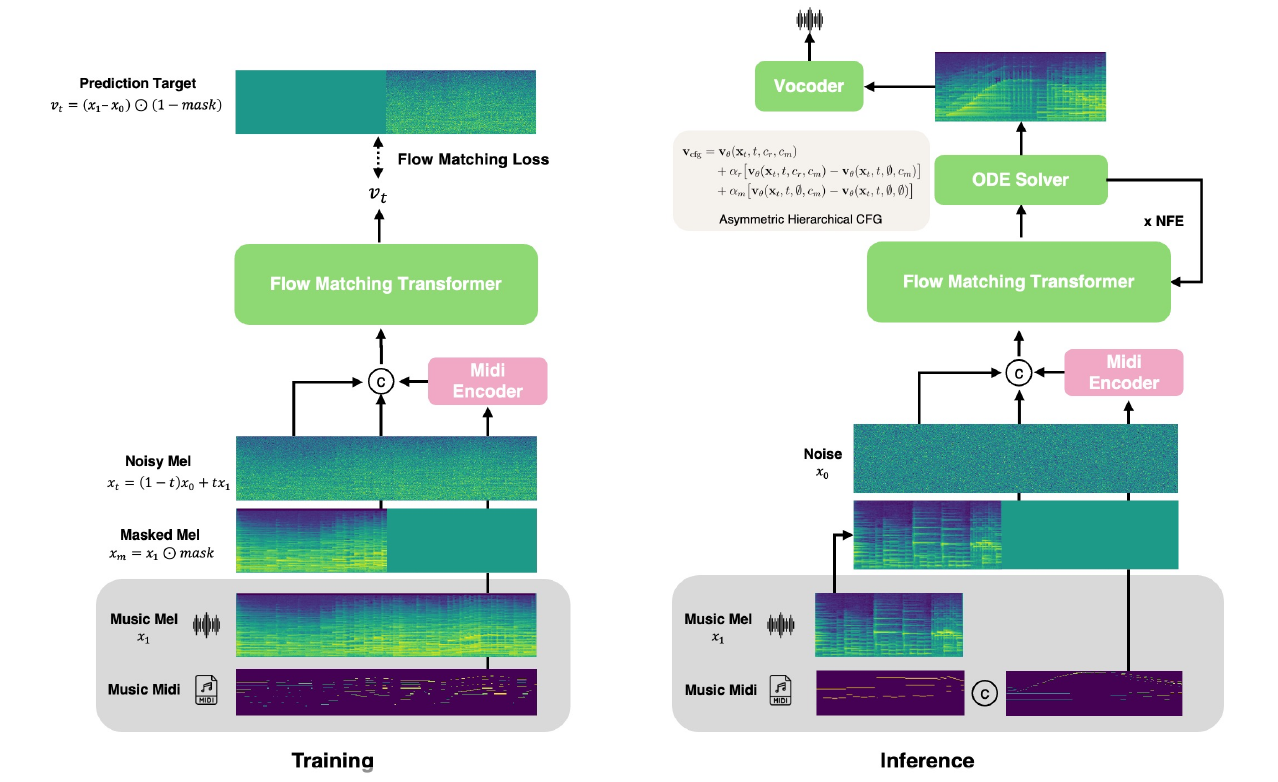}
  \caption{Overview of the \tool\ model architecture. The model performs in-context flow matching, conditioning on both \texttt{[Reference Audio, Reference MIDI]} and the \texttt{[Target MIDI]} to generate a mel spectrogram that is subsequently decoded by a vocoder. During inference, we apply an \textit{Asymmetric Hierarchical CFG} strategy to dynamically balance MIDI (semantic) guidance and reference-timbre (acoustic) guidance along the generation process.}
  \label{fig:model}
  \vspace{-10pt}
\end{figure*}

To ensure fine-grained controllability during inference, \tool further introduces an \textit{Asymmetric Hierarchical CFG} strategy. Standard compositional classifier-free guidance~\cite{liu2022compositional} symmetrically combines multiple conditions, which often causes gradient conflicts when dealing with tightly entangled musical attributes like pitch and timbre. By factoring the joint probability via the chain rule, our asymmetric formulation treats the target MIDI as a structural semantic anchor, while treating the reference prompt as a conditional acoustic refinement~\cite{brooks2023instructpix2pix}. This design naturally reflects the semantic-acoustic dependency of music generation, allowing the model to apply strong constant guidance scales without the two directions interfering with each other. 

Our main contributions are summarized as follows:
\begin{itemize}[leftmargin=*]
\item We propose \tool, an embedding-free, in-context flow matching synthesizer that bypasses the acoustic detail loss inherent in traditional retrieval-based embeddings. Experiments show \tool significantly outperforms token-based and diffusion-based baselines in audio quality, timbre similarity, and melody adherence.
\item We empirically demonstrate that zero-shot instrument cloning exhibits \textit{prompt-length scaling} under in-context learning. Unlike embedding-based systems, \tool consistently improves in both timbre fidelity and note accuracy as the reference prompt lengthens.
\item We introduce \textit{Asymmetric Hierarchical CFG}, a plug-and-play design, which formulates the inherent semantic-acoustic hierarchical nature of music and improve the overall generation quality. 

\end{itemize}

\section{\tool Framework}

\subsection{Task Formulation}

Let $\mathbf{x}^{\mathrm{ref}}$ be a short reference audio clip of $P$ frames, $\mathbf{m}^{\mathrm{ref}} \in \mathbb{R}^{P \times K}$ its time-aligned MIDI piano roll ($K{=}128$ pitch bins), and $\mathbf{m}^{\mathrm{tar}} \in \mathbb{R}^{L \times K}$ a target MIDI sequence of $L$ frames to be rendered. Zero-shot instrument cloning asks for target audio $\hat{\mathbf{x}}^{\mathrm{tar}}$ that (i)~follows the pitch, timing, and dynamics prescribed by $\mathbf{m}^{\mathrm{tar}}$, and (ii)~preserves the acoustic identity observed in $\mathbf{x}^{\mathrm{ref}}$, even for instruments \emph{unseen} during training. The formulation is agnostic to the internal audio representation: $\mathbf{x}$ may be a raw waveform, a time--frequency representation, or a pretrained latent feature, depending on the backbone. For \tool, we instantiate $\mathbf{x}$ as a normalized mel spectrogram with $D$ bins, and concatenate the reference and target along the time axis into a single in-context sequence $\mathbf{x} \in \mathbb{R}^{T \times D}$ and $\mathbf{m} \in \mathbb{R}^{T \times K}$ with $T = P + L$, supervising only the target segment during training.

\subsection{\tool Architecture}

Figure~\ref{fig:model} gives an overview of \tool. The model casts zero-shot instrument cloning as an in-context conditional flow matching problem: the reference audio--MIDI pair serves as an in-context prompt, and a Diffusion Transformer (DiT~\cite{peebles2023scalable}) learns to generate the target mel spectrogram conditioned on both the prompt (reference mel and MIDI) and the target MIDI. The generated spectrogram is decoded to a waveform by a pre-trained Vocos vocoder~\cite{siuzdak2024vocos}.

\subsubsection{In-Context Conditional Flow Matching}\label{sec:cfm}


\textbf{MIDI encoding.}
Typical MIDI piano roll only consists of a velocity-weighted note channel $\mathbf{r}_n \in \mathbb{R}^{T \times K}$, which often merges rapidly repeated pitches into a single sustained event, leading to note omissions. To deal with this issue, we additionally design a binary onset channel $\mathbf{r}_o \in \mathbb{R}^{T \times K}$ that explicitly indicates all note onset events.  
Each channel is projected to a hidden dimension $d_m$, and the resulting embeddings ($\mathbf{e}_n$ and $\mathbf{e}_o$) are fused via multiplicative gating:
\begin{equation}
  \mathbf{e}_m = \mathbf{e}_n + \mathbf{e}_o \odot \mathbf{e}_n,
  \label{eq:midi_enc}
\end{equation}
This allows onset events to sharpen note boundaries without overwriting the sustained pitch and dynamics carried by $\mathbf{e}_n$, thus representing the target note sequence more holistically.

\textbf{In-context reference conditioning.}
Instead of compressing the reference into a fixed-length embedding, \tool feeds the reference mel spectrogram directly as an in-context prefix. We construct $\tilde{\mathbf{x}}^{\mathrm{ref}} \in \mathbb{R}^{T \times D}$ by placing the real reference spectrogram in the first $P$ frames and zero-padding the remaining $L$ target frames. This representation enters the network alongside the noisy latent $\mathbf{x}_t$ and the MIDI embedding $\mathbf{e}_m$; the three are concatenated and linearly projected into the DiT hidden space. Because $\tilde{\mathbf{x}}^{\mathrm{ref}}$ lives in the same continuous space as the generation target, subsequent self-attention can freely attend across reference and target positions, retrieving fine-grained acoustic details---such as transients, formants, and room acoustics---without any information bottleneck. Meanwhile, the aligned reference MIDI explicitly teaches the model \emph{where} each acoustic cue belongs.

\textbf{Flow matching objective.}
Given a target mel spectrogram $\mathbf{x}$ and Gaussian noise $\mathbf{z} \sim \mathcal{N}(\mathbf{0}, \mathbf{I})$, we define the standard optimal transport flow matching path $\mathbf{x}_t = (1 - (1{-}\sigma)t)\mathbf{z} + t\mathbf{x}$ and its target velocity $\mathbf{v}^{*} = \mathbf{x} - (1{-}\sigma)\mathbf{z}$ (where $\sigma \to 0$). The network $\mathbf{v}_\theta$ is trained to predict this velocity via a masked $\ell_1$ loss:
\begin{equation}
  \mathcal{L} = \mathbb{E}_{t, \mathbf{x}, \mathbf{z}} \Big[\big\|\bigl(\mathbf{v}_\theta(\mathbf{x}_t,\, t,\, c_r,\, c_m) - \mathbf{v}^{*}\bigr) \odot \mathbf{M}\big\|_1\Big],
  \label{eq:loss}
\end{equation}
where $c_r$ and $c_m$ are the reference-timbre ($\mathbf{x}^{\mathrm{ref}}$) and MIDI conditions ($\mathbf{e}_m$), respectively. The mask $\mathbf{M}$ zeroes out both the reference-prompt frames and padding, ensuring that supervision is applied \emph{only} to the target segment.

\subsubsection{DiT Backbone}\label{sec:arch}

The backbone is a Diffusion Transformer (DiT)~\cite{peebles2023scalable}. Following standard DiT design, each transformer block employs adaptive layer normalization (AdaLN) to inject the diffusion timestep $t$ into both the self-attention and feed-forward sub-layers via predicted scale, shift, and gating parameters. To better capture the temporal dynamics of audio, we augment the standard architecture by applying a convolutional position embedding at the input layer for local temporal bias, as in Voicebox and F5-TTS~\cite{chen2025f5, le2023voicebox} and utilizing rotary position embeddings (RoPE)~\cite{su2024roformer} within the self-attention mechanism. A final AdaLN layer and linear projection map the hidden states back to the mel spectrogram dimension $D$.

\subsubsection{Asymmetric Hierarchical CFG}\label{sec:cfg}

\tool decouples MIDI and reference-timbre guidance at inference through an \emph{asymmetric hierarchical} classifier-free guidance (CFG) strategy. Here we take $c_r$ as the reference-timbre condition and $c_m$ as the MIDI condition.

Standard compositional diffusion~\cite{liu2022compositional} factorizes the joint posterior under a conditional-independence assumption, $p(c_m, c_r \mid \mathbf{x}_t) \approx p(c_m\mid \mathbf{x}_t)\,p(c_r\mid \mathbf{x}_t)$, which yields a symmetric additive combination of the two guidance directions and supports dual-control systems such as Stemgen~\cite{parker2024stemgen}. However, acoustic timbre ($c_r$) is inherently tied to the musical notes ($c_m$) being played. Thus, we adopt an asymmetric formulation inspired by the dual-conditioning strategy in InstructPix2Pix~\cite{brooks2023instructpix2pix}. By applying the chain rule, we factorize the joint probability as $p(c_m, c_r \mid \mathbf{x}_t) \propto p(c_r \mid c_m, \mathbf{x}_t)\, p(c_m \mid \mathbf{x}_t)$. This expresses a natural semantic--acoustic hierarchy: generation is first anchored in the MIDI structure, and then refined with instrument-specific acoustic features.

Translating the score decomposition of this log-posterior into the continuous-time flow matching framework, we compute the guided velocity field $\mathbf{v}_{\text{cfg}}$ using independent scales $\alpha_m, \alpha_r \ge 0$:
\begin{align}
  \mathbf{v}_{\text{cfg}} &= \mathbf{v}_\theta(\mathbf{x}_t, t, c_r, c_m) \nonumber \\
  &\quad + \alpha_m \bigl[ \mathbf{v}_\theta(\mathbf{x}_t, t, \emptyset, c_m) - \mathbf{v}_\theta(\mathbf{x}_t, t, \emptyset, \emptyset) \bigr] \nonumber\\
  &\quad + \alpha_r \bigl[ \mathbf{v}_\theta(\mathbf{x}_t, t, c_r, c_m) - \mathbf{v}_\theta(\mathbf{x}_t, t, \emptyset, c_m) \bigr].
  \label{eq:hier_cfg}
\end{align}
In this formulation, the second term ($\alpha_m$) pulls the unconditional velocity toward the MIDI-compliant direction. The third term ($\alpha_r$) then applies timbre guidance \emph{within} that MIDI context. 
we set $\alpha_m > \alpha_r$ considering that MIDI condition provides richer structural semantic information compared to the reference spectrogram.

\begin{figure*}[htbp]
  \centering
  \includegraphics[width=0.85\linewidth]{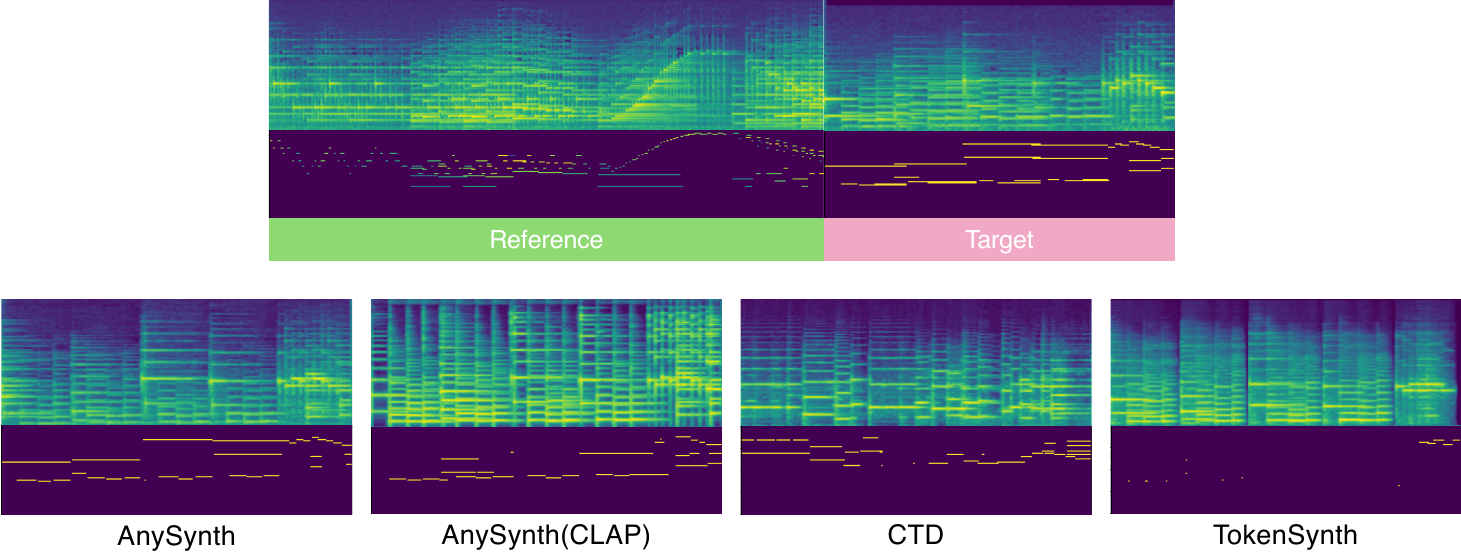}
  \caption{Spectrogram and transcribed MIDI visualizations. Top: The reference and target audio-MIDI pairs. Bottom: Model predictions from the same target MIDI conditioning.}
  \label{fig:casestudy}
  \vspace{-10pt}
\end{figure*}


\subsubsection{Inference}\label{sec:infer}

As illustrated in Figure \ref{fig:model}, at inference time, we leverage reference MIDI and Mel spectrograms alongside target MIDI sequences to perform instrument cloning. We sample $\mathbf{x}_0 \sim \mathcal{N}(\mathbf{0}, \mathbf{I})$ and integrate the hierarchically guided velocity field $\mathbf{v}_{\text{cfg}}$ (Eq.~\eqref{eq:hier_cfg}) using an Euler ODE solver with $N$ steps from $t{=}0$ to $t{=}1$. Crucially, each integration step requires only a single batched forward pass to compute the three velocity states required by Eq.~\eqref{eq:hier_cfg} in parallel. Thus, our asymmetric CFG adds no major computational overhead compared to standard CFG. After integration, we extract the target segment mel spectrogram, and decode it into an audio waveform using a pre-trained Vocos vocoder~\cite{siuzdak2024vocos}.

\section{Experiments}

\subsection{Experimental Setup}

\subsubsection{Implementation Details}

\noindent\textbf{Training Datasets}
We construct our training corpus by combining and rendering data from two primary sources: NSynth and Slakh. Here we consider three primary instrument classes--keyboard, guitar and bass.
\begin{itemize}[leftmargin=*, nosep]
    \item \textbf{NSynth}~\cite{engel2017neuralaudiosynthesismusical}: The Neural Audio Synthesis (NSynth) dataset is a large-scale, high-quality collection of over 300,000 annotated isolated musical notes. For our pipeline, the original 16 kHz note audio samples are first upsampled to 48 kHz using the LavaSR~\footnote{\url{https://huggingface.co/YatharthS/LavaSR}}. We then filter the dataset to include instruments from the `keyboard', `guitar', and `bass' families containing `acoustic' or `electronic' tags. The corresponding MIDI files are extracted from the Lakh dataset, grouped, and separated according to General MIDI (GM) rules. These are rendered using a rendering script, selecting 250 MIDI files per instrument.
    \item \textbf{Slakh}~\cite{manilow2019cuttingmusicsourceseparation}: The Synthesized Lakh (Slakh) dataset is a comprehensive multi-track audio dataset created by rendering the Lakh MIDI Dataset using professional-grade software synthesizers. We similarly filter the instruments specifically for the `keyboard', `guitar', and `bass'.
\end{itemize}

After merging these datasets, we perform an 8:2 train-test split based on distinct timbres. The training samples were constructed with prompt lengths ranging from 3 to 15 seconds and target generation lengths between 1.5 and 30 seconds. We ensured a minimum of 20 hours of data per timbre, yielding a total of 4,760 hours of training data.

\noindent\textbf{Training Configuration}
Our model utilizes a Flow Matching Transformer architecture. The backbone is a Diffusion Transformer (DiT)~\cite{peebles2023scalable} with a hidden dimension of 1024, a depth of 25 layers, and 16 attention heads with Rotary Position Embedding\cite{su2024roformer}; in total 481.30M parameters. 
For CFG (Sec.~\ref{sec:cfg}) training, we drop reference mel and Midi independently with a rate of $0.3$.
The flow matching process utilizes a cosine time scheduler~\cite{du2024cosyvoice} with a latent dimension of 128. Training was conducted across 8 GPUs with a batch size of 10 per GPU. We utilized the Adam optimizer with a peak learning rate of 1e-4. The learning rate schedule included a 32,000-step warmup period, followed by a linear decay to 0 over a total of 200,000 training steps.

\subsubsection{Evaluation Setup}

\noindent\textbf{Evaluation Datasets}
In addition to the test splits of the NSynth and Slakh datasets, our evaluation incorporates two real-recorded dataset, namely MAESTRO and GuitarSet~\cite{manilow2019cuttingmusicsourceseparation, xi2018guitarset}. Specifically, MAESTRO comprises approximately 200 hours of annotated piano recordings, while GuitarSet consists of 6 hours of live guitar performances.
To ensure a fair comparison with our baseline models which are constrained to generating a maximum of 5 seconds of audio, we segmented our test set accordingly into 5-second target clips. For each evaluated timbre, we provide conditioning prompts of 3, 8, and 15 seconds in length, extracting 50 segments per timbre. The final curated testset contains approximately 2 hours of audio.
\begin{figure*}[htpb]
  \centering
  \includegraphics[width=0.8\textwidth]{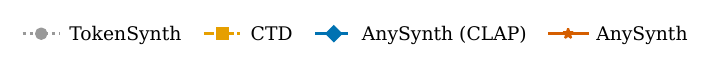}\\[2pt]
  \begin{subfigure}[b]{0.32\textwidth}
    \includegraphics[width=\linewidth]{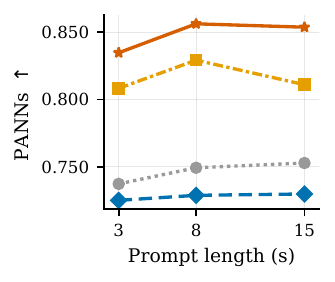}
    \caption{PANNs}
  \end{subfigure}\hfill
  \begin{subfigure}[b]{0.32\textwidth}
    \includegraphics[width=\linewidth]{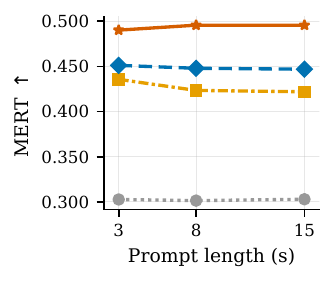}
    \caption{MERT}
  \end{subfigure}\hfill
  \begin{subfigure}[b]{0.32\textwidth}
    \includegraphics[width=\linewidth]{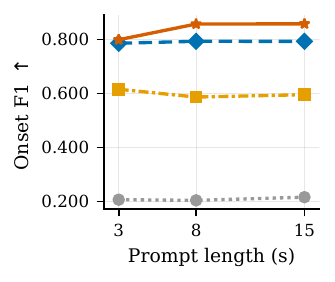}
    \caption{Onset F1}
  \end{subfigure}
  \caption{Prompt-length scaling results of all models averaged over the four evaluation datasets (NSynth, Slakh, MAESTRO, GuitarSet).}
  \label{fig:prompt_scaling_icl}
  \vspace{-15pt}
\end{figure*}

\begin{table}[!htpb]
  \centering
  \caption{Objective evaluation of zero-shot instrument cloning across multiple datasets, reporting mean results averaged over three reference prompt lengths (3s, 8s, 15s). The Ground Truth row reports dataset-level upper-bound references derived from real recordings.}
  \label{tab:main_results}
  \footnotesize
  \setlength{\tabcolsep}{3pt}
  \begin{tabular}{l ccccc}
  \toprule
  \textbf{Model} & \textbf{MERT}$\uparrow$ & \textbf{PANNs}$\uparrow$ & \textbf{Onset F1}$\uparrow$ & \textbf{CE}$\uparrow$ & \textbf{PQ}$\uparrow$ \\
  \midrule

  \rowcolor{gray!20}
  \multicolumn{6}{c}{\textbf{NSynth}} \\
  \midrule
  Ground Truth & 1.000 & 0.854 & 0.805 & 4.980 & 6.799 \\
  \cmidrule{1-6}
  TokenSynth~\cite{kim2025tokensynth} & 0.313 & 0.781 & 0.190 & 4.880 & 6.907 \\
  CTD~\cite{demerle2024combining} & 0.399 & 0.779 & 0.503 & \textbf{5.783} & 6.897 \\
  \tool (CLAP) & 0.474 & 0.765 & \textbf{0.709} & 5.521 & 7.002 \\
  \tool & \textbf{0.513} & \textbf{0.820} & 0.704 & 5.730 & \textbf{7.175} \\
  \midrule

  \rowcolor{gray!20}
  \multicolumn{6}{c}{\textbf{Slakh}} \\
  \midrule
  Ground Truth & 1.000 & 0.867 & 0.947 & 5.351 & 7.173 \\
  \cmidrule{1-6}
  TokenSynth~\cite{kim2025tokensynth} & 0.292 & 0.769 & 0.214 & 4.696 & 6.802 \\
  CTD~\cite{demerle2024combining} & 0.491 & 0.869 & 0.625 & 5.221 & 6.734 \\
  \tool (CLAP) & 0.428 & 0.730 & 0.808 & 4.853 & 6.766 \\
  \tool & \textbf{0.519} & \textbf{0.882} & \textbf{0.853} & \textbf{5.324} & \textbf{7.146} \\
  \midrule

  \rowcolor{gray!20}
  \multicolumn{6}{c}{\textbf{MAESTRO}} \\
  \midrule
  Ground Truth & 1.000 & 0.877 & 0.968 & 6.914 & 7.257 \\
  \cmidrule{1-6}
  TokenSynth~\cite{kim2025tokensynth} & 0.297 & 0.779 & 0.129 & 5.891 & 6.957 \\
  CTD~\cite{demerle2024combining} & 0.393 & 0.813 & 0.491 & 6.186 & 6.194 \\
  \tool (CLAP) & 0.461 & 0.771 & 0.765 & 6.195 & 7.451 \\
  \tool & \textbf{0.484} & \textbf{0.885} & \textbf{0.894} & \textbf{6.863} & \textbf{7.459} \\
  \midrule

  \rowcolor{gray!20}
  \multicolumn{6}{c}{\textbf{GuitarSet}} \\
  \midrule
  Ground Truth & 1.000 & 0.928 & 0.954 & 7.433 & 8.129 \\
  \cmidrule{1-6}
  TokenSynth~\cite{kim2025tokensynth} & 0.308 & 0.656 & 0.303 & 5.411 & 7.447 \\
  CTD~\cite{demerle2024combining} & 0.425 & 0.804 & 0.779 & \textbf{6.321} & 7.800 \\
  \tool (CLAP) & 0.432 & 0.646 & 0.881 & 5.262 & 7.046 \\
  \tool & \textbf{0.460} & \textbf{0.805} & \textbf{0.959} & 6.135 & \textbf{7.995} \\
  \bottomrule
  \end{tabular}
\vspace{-10pt}
\end{table}

\noindent\textbf{Evaluation Metrics}
\begin{itemize}[leftmargin=*, nosep]
    \item \textbf{PANNs}: We measure timbre consistency by computing the cosine similarity between the PANNs embeddings~\cite{kong2020panns} of the generated audio and a 15-second reference prompt, sharing the same timbre.
    
    \item \textbf{Onset F1}: We assess the structural accuracy of the generated melody by calculating the MIDI-level Onset F1 Score. The MIDIs are transcribed from the generated audio using YourMT3+~\cite{chang2024yourmt3+}.

    \item \textbf{MERT}: To comprehensively evaluate both timbre fidelity and melodic accuracy, we calculate the patch-level cosine similarity between the Mert Featurel~\cite{li2024mert}\footnote{\url{https://huggingface.co/m-a-p/MERT-v1-95M}} of the generated audio and the ground truth (GT).
    
    \item \textbf{Audiobox Aesthetics}: We evaluate the overall perceptual and acoustic quality of the generations using the Content Enjoyment (CE) and Production Quality (PQ) metrics derived from the Audiobox Aesthetics evaluator~\cite{tjandra2025meta}.
    
\end{itemize}

\noindent\textbf{Baselines}
We compare against representative embedding-based and token-based neural synthesizers, including TokenSynth~\cite{kim2025tokensynth}, Control-Transfer-Diffusion (CTD)~\cite{demerle2024combining}, and an \tool\ variant that replaces in-context reference conditioning with CLAP embeddings. The \tool\ (CLAP) comparison is designed to isolate the benefit of in-context conditioning from the benefit of the DiT backbone itself.

\begin{figure*}[htpb]
  \centering
  \includegraphics[width=0.9\textwidth]{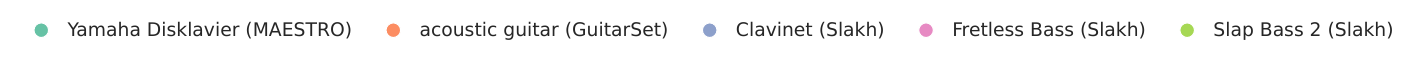}\\[2pt]
  \begin{subfigure}[b]{0.19\textwidth}
    \includegraphics[width=\linewidth]{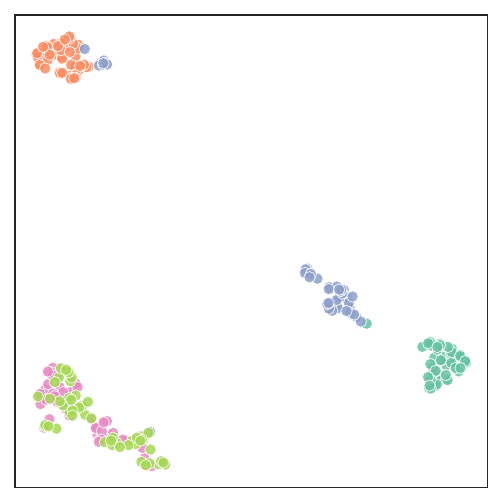}
    \caption{Ground Truth}
  \end{subfigure}\hfill
  \begin{subfigure}[b]{0.19\textwidth}
    \includegraphics[width=\linewidth]{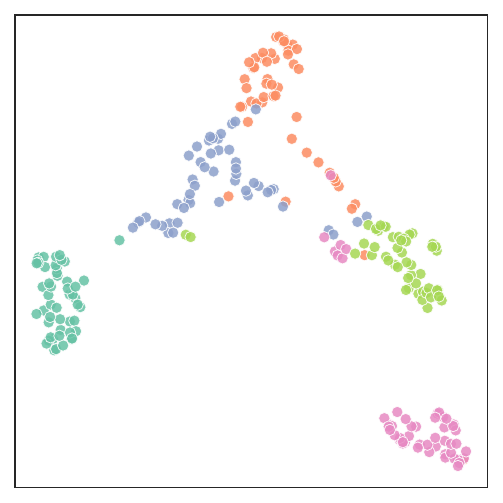}
    \caption{AnySynth}
  \end{subfigure}\hfill
  \begin{subfigure}[b]{0.19\textwidth}
    \includegraphics[width=\linewidth]{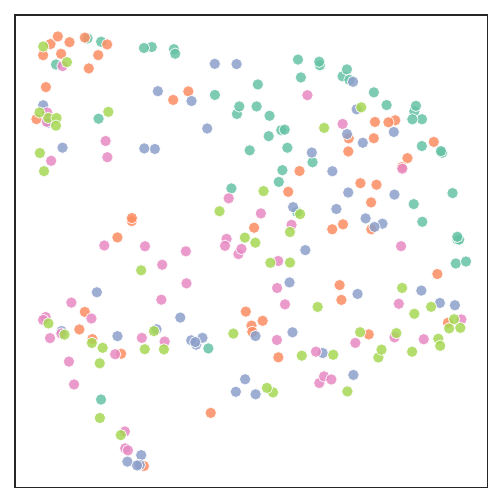}
    \caption{AnySynth (CLAP)}
  \end{subfigure}\hfill
  \begin{subfigure}[b]{0.19\textwidth}
    \includegraphics[width=\linewidth]{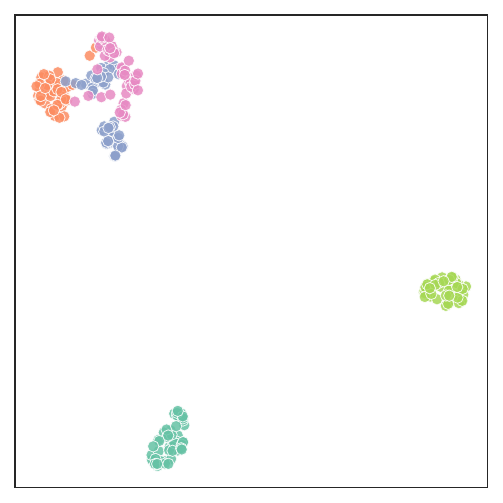}
    \caption{CTD}
  \end{subfigure}\hfill
  \begin{subfigure}[b]{0.19\textwidth}
    \includegraphics[width=\linewidth]{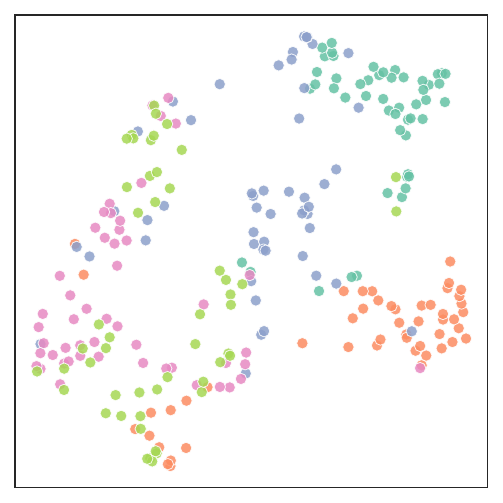}
    \caption{TokenSynth}
  \end{subfigure}
  \caption{UMAP projections of PANNs embeddings, computed independently per model on the same set of samples drawn from five instruments across MAESTRO, GuitarSet, and Slakh.}
  \label{fig:umap_intra_model}
  \vspace{-5pt}
\end{figure*}

\subsection{Main Results}

Table~\ref{tab:main_results} summarizes the quantitative comparison averaged over 3\,s / 8\,s / 15\,s reference prompts. Across the four datasets, \tool\ ranks overall first on every metric, confirming that \tool simultaneously improves timbre fidelity, melody adherence, and perceptual quality. 
Notably, it demonstrates robust generalization capabilities on real-world recordings, successfully overcoming the degradation issues prevalent in previous methods. 

\noindent\textbf{Qualitative evidence.} Figure~\ref{fig:umap_intra_model} visualizes UMAP projections\cite{mcinnes2018umap} of PANNs embeddings for samples from five instruments across 5 instruments from MAESTRO, GuitarSet, and Slakh. Consistent with the quantitative findings in Table~\ref{tab:main_results}, models achieving higher PANNs scores exhibit well-defined clustering and clear separation between timbres, whereas models with lower scores fail.
Figure~\ref{fig:casestudy} further shows that \tool\ recovers fine spectral transients and vibrato from the reference and the transcribed MIDI still aligns with the target, while baselines either blur the spectrogram or deviate from the target melody.

\subsection{Ablation Studies}
\begin{table}[!htpb]
  \centering
  \caption{Objective evaluation of zero-shot instrument cloning across different prompt length. The results demonstrate that our proposed model, \tool, consistently outperforms baseline methods across all metrics.}
  \label{tab:prompt_scaling}
  \footnotesize
  \setlength{\tabcolsep}{6pt}
  \begin{tabular}{l ccc}
  \toprule
  \textbf{Model} & \textbf{MERT}$\uparrow$ & \textbf{PANNs}$\uparrow$ & \textbf{Onset F1}$\uparrow$ \\
  \midrule

  \rowcolor{gray!20}
  \multicolumn{4}{c}{\textbf{Prompt Length: 3s}} \\
  \midrule
  TokenSynth~\cite{kim2025tokensynth} & 0.303 & 0.737 & 0.207 \\
  CTD~\cite{demerle2024combining}     & 0.436 & 0.808 & 0.616 \\
  AnySynth (CLAP)                     & 0.451 & 0.725 & 0.786 \\
  AnySynth                            & \textbf{0.490} & \textbf{0.835} & \textbf{0.842} \\
  \midrule

  \rowcolor{gray!20}
  \multicolumn{4}{c}{\textbf{Prompt Length: 8s}} \\
  \midrule
  TokenSynth~\cite{kim2025tokensynth} & 0.302 & 0.749 & 0.204 \\
  CTD~\cite{demerle2024combining}     & 0.424 & 0.829 & 0.587 \\
  AnySynth (CLAP)                     & 0.448 & 0.729 & 0.793 \\
  AnySynth                            & \textbf{0.496} & \textbf{0.856} & \textbf{0.857} \\
  \midrule

  \rowcolor{gray!20}
  \multicolumn{4}{c}{\textbf{Prompt Length: 15s}} \\
  \midrule
  TokenSynth~\cite{kim2025tokensynth} & 0.303 & 0.753 & 0.216 \\
  CTD~\cite{demerle2024combining}     & 0.422 & 0.811 & 0.596 \\
  AnySynth (CLAP)                     & 0.447 & 0.730 & 0.793 \\
  AnySynth                            & \textbf{0.496} & \textbf{0.854} & \textbf{0.858} \\
  \bottomrule
  \end{tabular}
  \vspace{-15pt}
\end{table}

\subsubsection{In-Context Conditioning vs. CLAP Conditioning}\label{sec:ablation_icl}
Table~\ref{tab:main_results} isolates the impact of our conditioning strategy by comparing \tool\ against a baseline using a fixed $512$-d CLAP embedding (\tool-CLAP). Replacing the in-context prompt with CLAP consistently degrades timbre similarity, with PANNs dropping by $5.3\%$ to $23.3\%$ across datasets, and MERT dropping by up to $17.4\%$. The degradation is most severe on GuitarSet and Slakh, demonstrating that global embeddings compress away the fine transients and artifacts essential for realistic timbre reproduction. Overall, this ablation confirms that \tool's improvements stem primarily from the in-context conditioning mechanism rather than the DiT backbone.

\begin{table}[htpb]
  \centering
  \caption{Ablation on CFG formulations. All variants share identical weights and sampling steps. Scales denote a single numerical value $\alpha$ or a tuple $(\alpha_m, \alpha_r)$.}
  \footnotesize
  \setlength{\tabcolsep}{3pt}
  \label{tab:ablation_cfg}
  \begin{tabular}{lccc}
    \toprule
    \textbf{CFG/Scales} & \textbf{MERT}$\uparrow$ & \textbf{PANNs}$\uparrow$ & \textbf{Onset F1}$\uparrow$ \\
    \midrule
    Eq.~\eqref{eq:combined_cfg} / $2.0$&0.492&0.845&0.839\\
    Eq.~\eqref{eq:stemgen_cfg} / $(2.0, 2.0)$&0.465&0.796&0.649\\
    Eq.~\eqref{eq:stemgen_cfg} / $(2.0, 1.0)$&0.465& 0.779&0.655\\
    Eq.~\eqref{eq:hier_cfg} / $(2.0, 1.0)$(ours)&\textbf{0.494}&\textbf{0.848}&\textbf{0.852} \\
    \bottomrule
  \end{tabular}
  \vspace{-10pt}
\end{table}



\subsubsection{Prompt Length Scaling}\label{ablation_prompt_length}

Figure~\ref{fig:prompt_scaling_icl} and Table~\ref{tab:prompt_scaling} report PANNs, MERT, and Onset F1 related to the reference-prompt length, averaged over the four evaluation datasets. 

As the prompt length increases from 3\,s to 15\,s, \tool\ exhibits a clear upward trend across all metrics. In contrast, the baseline methods do not demonstrate consistent benefits from longer context.

We observe no significant performance gain when increasing the prompt duration from 8 s to 15 s. This observation suggests two potential explanations: First, an 8-second prompt may already provide sufficient acoustic information, leading to diminishing marginal returns with further extension. Second, the relative simplicity of generating a 5-second segment may not fully leverage the benefits of longer contexts. We hypothesize that the advantages of extended prompts would become more pronounced in the synthesis of longer audio sequences, which need to be confirmed by future works.


\subsubsection{Asymmetric Hierarchical CFG}\label{sec:ablation_cfg}

We compare three CFG formulation varients with the same checkpoint of \tool. 

The variants are as follows:

\textbf{(i) Combined CFG.} Treats $(c_r, c_m)$ as a single composite condition, using one scalar $\alpha$ to amplify the fully-conditioned direction:
\begin{align}
  \scriptsize
  \mathbf{v}_{\text{cfg}}^{\mathrm{comb}} &= \mathbf{v}_\theta(\mathbf{x}_t, t, c_r, c_m) \nonumber \\
  &\quad + \alpha \bigl[ \mathbf{v}_\theta(\mathbf{x}_t, t, c_r, c_m) - \mathbf{v}_\theta(\mathbf{x}_t, t, \emptyset, \emptyset) \bigr].
  \label{eq:combined_cfg}
\end{align}

\textbf{(ii) Symmetric CFG.} Assumes $c_m$ and $c_r$ are independent, additively combining their marginal guidances from the unconditional baseline:
\begin{align}
  \mathbf{v}_{\text{cfg}}^{\mathrm{ind}} &= \mathbf{v}_\theta(\mathbf{x}_t, t, \emptyset, \emptyset) \nonumber \\
  &\quad + \alpha_m \bigl[ \mathbf{v}_\theta(\mathbf{x}_t, t, \emptyset, c_m) - \mathbf{v}_\theta(\mathbf{x}_t, t, \emptyset, \emptyset) \bigr] \nonumber \\
  &\quad + \alpha_r \bigl[ \mathbf{v}_\theta(\mathbf{x}_t, t, c_r, \emptyset) - \mathbf{v}_\theta(\mathbf{x}_t, t, \emptyset, \emptyset) \bigr].
  \label{eq:stemgen_cfg}
\end{align}

\textbf{(iii) Asymmetric CFG (Ours).} Employs the hierarchical decomposition defined in Eq.~\eqref{eq:hier_cfg}.

We evaluate these variants on the four evaluation datasets with 3\,s, 8\,s, 15\,s reference prompt and compute the average value among all datasets and prompt length. As shown in Table~\ref{tab:ablation_cfg}, we conclude two observations:
\begin{itemize}[leftmargin=*, nosep]
    \item \textbf{Invalidity of the independence assumption:} Symmetric CFG assumes that the MIDI ($c_m$) and reference ($c_r$) conditions are strictly independent. However, comparing Symmetric CFG (Eq.~\eqref{eq:stemgen_cfg}/ $(\alpha_m, \alpha_r) = (2.0, 2.0), (2.0, 1.0)$) with Combined CFG (Eq.~\eqref{eq:combined_cfg}/ $\alpha=2.0$) and Asymmetric CFG (Eq.~\eqref{eq:hier_cfg}/$(\alpha_m, \alpha_r) = (2.0, 1.0)$), we observe a significant decline in performance metrics. This highlights that the strict independence assumption is inappropriate for tightly entangled attributes like pitch and timbre, ultimately impairing the overall generation quality.
    \item \textbf{Implicit semantic-acoustic inductive bias:} While Combined CFG (Eq.~\eqref{eq:combined_cfg}/ $\alpha=2.0$) achieves highly competitive timbre similarity, our Asymmetric CFG demonstrates superior overall generation quality. This comprehensively validates our design: by explicitly formulating the target MIDI as the structural anchor and the reference prompt as the conditional refinement, Asymmetric CFG successfully embeds the well-established "semantic-then-acoustic" inductive bias directly into the DiT architecture.
\end{itemize}

\section{Conclusion}

In this paper, we introduced \tool, reformulating zero-shot instrument cloning as an in-context learning problem. By directly conditioning a Diffusion Transformer on uncompressed reference audio and target MIDI, \tool bypasses the semantic bottleneck of traditional embedding methods. Experiments demonstrate it significantly outperforms baselines in audio quality, timbre fidelity, and melody adherence. Notably, \tool exhibits \textit{prompt-length scaling}, consistently improving generation quality as the reference lengthens. Furthermore, our novel \textit{Asymmetric Hierarchical CFG} structurally achieve better audio quality, timbre similarity and melody adherence. Ultimately, \tool sets a new standard for expressive zero-shot instrument cloning.



\section{AI Usage Statement}

Generative AI tools are employed for language polishing, structural refinement of the paper writing, and assistance with routine coding tasks (e.g., standard utility scripts). All core scientific ideas, novel algorithmic designs, experimental methodologies, and data analyses are entirely the work of human. The authors have reviewed all AI-assisted text and code, and take full responsibility for the integrity, accuracy, and originality of the paper work.



\bibliography{ref}

\end{document}